\documentclass[preprint,superscriptaddress]{revtex4}

\usepackage{amssymb}

\usepackage{graphicx}
\usepackage{dcolumn}
\usepackage{bm}
\usepackage[CJKbookmarks=true, dvipdfm, colorlinks, linkcolor=blue, anchorcolor=black, citecolor=blue, urlcolor=black]{hyperref}

\begin{document}


\title{Microscopic coexistence of antiferromagnetic order and superconductivity in Ba$_{0.77}$K$_{0.23}$Fe$_{2}$As$_{2}$}

\author{Z. Li}
\affiliation{Institute of Physics and Beijing National Laboratory for Condensed Matter Physics, Chinese Academy of Sciences, Beijing 100190, China}

\author{R. Zhou}
\affiliation{Institute of Physics and Beijing National Laboratory for Condensed Matter Physics, Chinese Academy of Sciences, Beijing 100190, China}
\author{Y. Liu}
\affiliation{Max Planck Institute - Heisenbergstrasse 1, D-70569 Stuttgart, Germany}
\author{D. L. Sun}
\affiliation{Max Planck Institute - Heisenbergstrasse 1, D-70569 Stuttgart, Germany}
\author{J. Yang}
\affiliation{Institute of Physics and Beijing National Laboratory for Condensed Matter Physics, Chinese Academy of Sciences, Beijing 100190, China}
\author{C. T. Lin}
\affiliation{Max Planck Institute - Heisenbergstrasse 1, D-70569 Stuttgart, Germany}
\author{Guo-qing Zheng}%
\affiliation{Institute of Physics and Beijing National Laboratory for Condensed Matter Physics, Chinese Academy of Sciences, Beijing 100190, China}
\affiliation{Department of Physics, Okayama University, Okayama 700-8530, Japan}

\date{\today}

\begin{abstract}
We report $^{75}$As nuclear magnetic resonance studies on an underdoped single-crystal Ba$_{0.77}$K$_{0.23}$Fe$_{2}$As$_{2}$ with  $T_{\rm{c}}$=16.5 K. Below  $T_{\rm{N}}$=46 K, the NMR peaks for $H \parallel c$ split and those  for $H \parallel a$ shift to  higher frequencies, which indicates that an internal magnetic field along the $c$-axis develops below $T_{\rm{N}}$. The spin lattice relaxation rate $1/T_{1}$ measured at the shifted peak with $H \parallel a$ which experiences the internal field shows a distinct decrease below $T_{\rm{c}}(\mu_{0}H$=12 T) = 16 K, following a $T^3$ relation at low temperatures. Our results show unambiguously that antiferromagnetic order and  superconductivity coexist microscopically. The unusual superconducting state with the coexisting magnetism is highlighted.

\end{abstract}

\maketitle


Magnetism  and superconductivity (SC) are two outstanding  quantum phenomena,  and the relationship between the magnetic and superconducting orders has naturally become  an important subject  in condensed matter physics. It is well known that magnetism is harmful for conventional $s$-wave superconductivity. In the last decade or so,   whether
antiferromagnetism (AFM) and unconventional SC can coexist at a microscopic scale  has been one of the central issues.
In heavy fermion compounds, there is strong evidence that
AFM and SC coexist homogeneously and microscopically \cite{Kawasaki,Zheng,Sato}. In cuprate high transition-temperature ($T_{\rm{c}}$) superconductors, there are also indications that SC can coexist with AFM at a microscopic scale under certain circumstances \cite{Kotegawa}.

In the recently discovered  iron pnictides, superconductivity also emerges in the vicinity of antiferromagnetism \cite{YKamihara,XHChen}. Therefore, the relationship between  AFM and SC is of great importance for understanding the physics of this new class of superconductors. It has been proposed that elucidating such a relationship
can serve to determine the pairing symmetry, which is unsettled yet. It was shown that conventional $s^{++}$-wave SC is hard to coexist with AFM, while sign-change $s^{+-}$-wave SC can \cite{Schmalian}. Furthermore, this issue is directly related to possible quantum critical phenomena which is a widely studied subject in various classes of materials \cite{Sachdev}. A microscopic coexistence of AFM and SC is a necessary condition for a quantum critical point beneath the superconducting dome which is proposed to exist in cuprate high-$T_{\rm{c}}$ superconductors \cite{Broun}.

 Early experiments including nuclear magnetic resonance (NMR) measurement in the iron-pnictide superconductor Ba$_{1-x}$K$_{x}$Fe$_{2}$As$_{2}$ have suggested that, although AFM and SC occur in the same sample, the two phenomena take place at different, phase separated, parts of the sample \cite{Julien,JTPark}. Although there are also recent suggestions  that SC and AFM may coexist in  Ba$_2$Fe$_{2}$As$_{2}$ replaced by various elements such as Ca, K (to replace Ba) \cite{Baek,Urbano,Wiesenmayer,Avci}, Co (to replace Fe) \cite{Julien,Laplace},  or P (to replace As) \cite{Iye}, or in SmFeAsO$_{1-x}$F$_x$ \cite{Sanna},  the onset of the SC was only evidenced  by a susceptibility measurement, but not by a microscopic probe. For example, no sharp change in other physical quantities such as the nuclear spin-lattice relaxation rate is found {\it right} below $T_c$. Thus, the relationship between the AFM and SC in the iron-pnictides is still controversial due to lacks of a suitable experimental probe or a high quality sample. Therefore, a measurement using a single microscopic experimental technique capable of probing  both orders  in a high quality sample is highly desired.

Here we report $^{75}$As NMR measurements on an underdoped single-crystal Ba$_{0.77}$K$_{0.23}$Fe$_{2}$As$_{2}$ with  $T_{\rm{c}}$=16.5 K. Below  $T_{\rm{N}}$=46 K, the NMR transition peaks for $H \parallel c$ split and those  for $H \parallel a$ shift to  higher frequencies, which indicates that the antiferromagnetic order sets in, with the ordered Fe moment lying on the $ab$-plane. The spin lattice relaxation rate $1/T_{1}$ measured at the central transition peak with $H \parallel a$ shows distinct decreases at $T_{\rm{N}}$=46 K and $T_{\rm{c}}(\mu_{0}H$=12 T)=16 K, respectively. Since the nuclei corresponding to the shifted peak experience an internal magnetic field due to the electrons  in the antiferromagnetic ordered state below $T_{\rm{N}}$, our results show unambiguously that the electrons that are hyperfine coupled to the nuclei are responsible for both  antiferromagnetic order and the superconductivity. We also discuss the property of the superconducting state coexisting with magnetism.

The single crystals of Ba$_{1-x}$K$_{x}$Fe$_{2}$As$_{2}$ with $0.23 \leq x \leq 1$ were grown by using the self-flux method and the K content was determined by energy dispersive X-ray spectroscopy (EDX) \cite{GLSun}. The typical error for $x$  is less than $\pm$0.02.
The samples with $x$=0.23, 0.24, 0.32 \cite{Li} and 0.61  were selected for NMR measurements. Each sample has a typical surface size of 4 mm$\times$1.5 mm.
The $T_{\rm{c}}$ was measured by both DC susceptibility using a superconducting quantum interference device and by AC susceptibility using an \emph{in situ} NMR coil at zero field and at $\mu_{0}H$=12 T. For the $x$=0.23 sample ($T_{\rm{c}}$=16.5 K),   $T_{\rm{c}}$  decreases to 16 K for the $\mu_{0}H_{0} (=12$ T) $\parallel a$ axis and to $15.5$ K for the $\mu_{0}H_{0} (=12$ T) $\parallel c$ axis.
The spectra of $^{75}$As with the nuclear gyromagnetic ratio $\gamma=7.2919$ MHz/T are obtained by scanning the RF frequency at a fixed magnetic field $\mu_{0}H_{0}(=11.9977$ T). The  $1/T_{1}$ was determined from a good
fit of the nuclear magnetization to $1-M(t)/M(\infty)=0.1$exp$(-t/T_{1})+0.9$ exp$(-6 t/T_{1})$ for the central transition peak, where $M(t)$ is the nuclear magnetization at time $t$ after the saturation pulse \cite{ANarath}.

\begin{figure}
\includegraphics[width=8cm,clip]{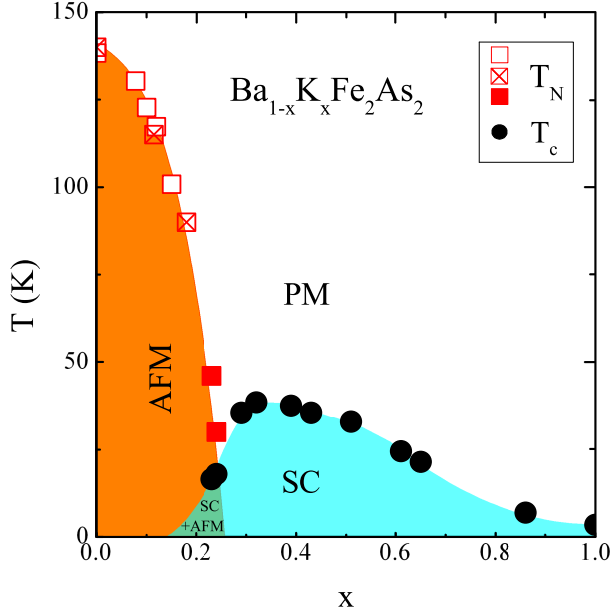}
\caption{\label{fig:Li_Fig1} (Color online) Phase diagram of Ba$_{1-x}$K$_{x}$Fe$_{2}$As$_{2}$. The solid squares designate the N\'{e}el temperature $T_{\rm{N}}$ determined by NMR measurements,  and the other data points for $T_{\rm{N}}$ are from Ref.\cite{MRotter,Wen}. The solid circles indicate superconducting transition temperature $T_{\rm{c}}$ determined from susceptibility measurements. AFM, PM and SC represent antiferromagnetically-ordered, paramagnetic and superconducting states, respectively.}
\end{figure}

Figure \ref{fig:Li_Fig1} shows the phase diagram of Ba$_{1-x}$K$_{x}$Fe$_{2}$As$_{2}$.
The solid squares designate the N\'{e}el temperature $T_{\rm{N}}$ determined from NMR spectra and $T_{1}$ measurements (see below), the data points shown by symbol $\boxtimes$ are adapted from Ref.\cite{MRotter} for poly-crystals, and the open squares are  from Ref.\cite{Wen} for single crystals. $T_N$ for our $x$=0 sample determined from resistivity \cite{GLSun} agrees well with Ref.\cite{Wen}.
The solid circles indicate  $T_{\rm{c}}$ determined from susceptibility measurements. The samples with $x$=0.23 and 0.24 belong to the underdoped regime.

\begin{figure}
\includegraphics[width=9cm,clip]{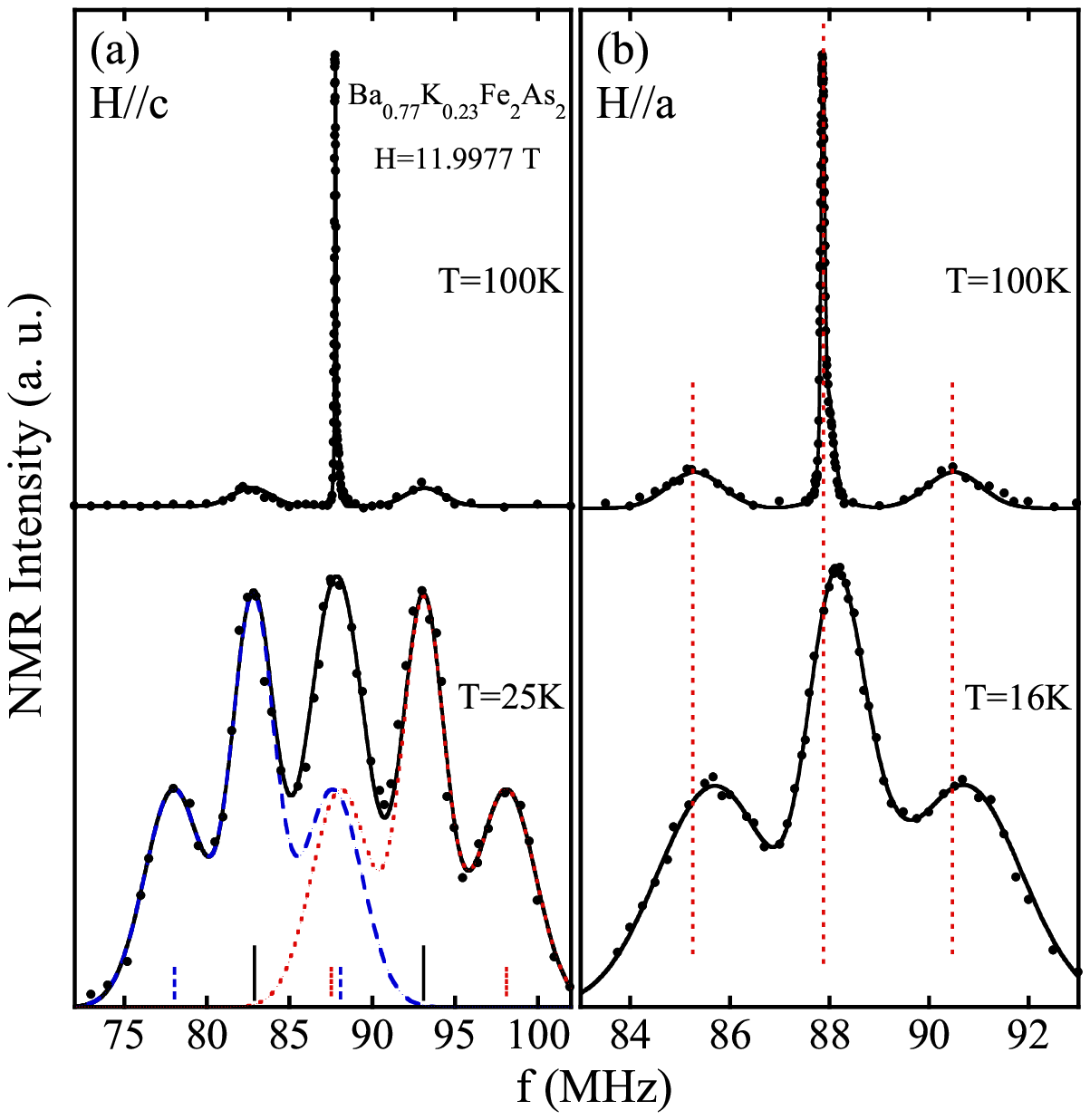}
\caption{\label{fig:Li_Fig2} (Color online) $^{75}$As NMR spectra above and below $T_{\rm{N}}$ for (a) $H \parallel c$ and (b) $H \parallel a$ respectively. The vertical axis for the  $T=100$ K spectra was offset for clarity. (a) The blue dashed curve and the red dotted curve are the simulated two sets of spectra that are split by the internal magnetic field,  and the black curve is the sum of the  two sets of spectra. The short lines designate peak positions. (b) The straight dashed lines designate peak positions of the $T=100$ K spectrum. }
\end{figure}

Figure \ref{fig:Li_Fig2} shows the frequency-scanned $^{75}$As NMR spectra  for Ba$_{0.77}$K$_{0.23}$Fe$_{2}$As$_{2}$ with both $H \parallel c$-axis and $H \parallel a$-axis configurations. The spectrum at $T=100$ K  in the paramagnetic state shows a sharp central peak accompanied by two satellite peaks due to nuclear quadrupole interaction. The nuclear quadrupole frequency $\nu_{\rm{Q}}$ is found to be $5.06$ MHz, which is a little smaller than the optimal doped sample ($5.1$ MHz) \cite{Li}. The spectra change below $T_{\rm{N}}$=46 K. Namely, all the three peaks split for $H \parallel c$, while the peaks shifted to  higher frequencies for $H \parallel a$. In the antiferromagnetically ordered state, neutron experiments have found that Fe magnetic moments lie on the $ab$ plane, forming stripes\cite{WBao}. The internal magnetic field  at the As site located above or below the magnetically ordered Fe layer should direct along the $c$ axis or anti-parallel to the $c$ axis. In such a case, for $H \parallel c$, the effective field seen by the As nuclei sitting above or below the Fe layer is $H^{\rm eff}_{\rm{c}}=H_{0} \pm H_{\rm{int}}$, which will split the spectra into two sets. One set consisting of the central transition and two satellites shifted up by the amount of $\gamma H_{\rm int}$,  which corresponds to the As sitting above the Fe layer, and the other set  corresponding to the As sitting below the Fe layer shifted down by the same amount. For $H \parallel a$,  on the other hand, $H^{\rm eff}_{\rm{a}}=\sqrt{H_{0}^{2}+H_{\rm{int}}^{2}}$ will simply shift each peak to a higher resonance frequency \cite{KKitagawa,Supple}.
The spectra shown in Fig. \ref{fig:Li_Fig2}  show that the As nuclei indeed experience such  internal magnetic fields  below $T_{\rm{N}}$=46 K. The same is true for the $x$=0.24 sample (data not shown).

As seen in Fig. \ref{fig:Li_Fig2} (a), for $H \parallel c$,  the two sets of the spectra  happen to overlap with each other, resulting in five peaks, of which the central one is the broadest. The solid curve is the simulation of the summation of the  two sets of the spectra. In the calculation, the area ratio of a satellite peak to the central peak is set to 3:4  according to the theoretical requirement which is indeed satisfied at $T$=100 K. Such calculation fits  the spectra very well below $T$=35 K, indicating that the whole sample is in the antiferromagnetically ordered state below this temperature. However,  in the temperature range between 35 K and $T_{\rm{N}}$=46 K, the agreement between the calculation and the observed spectra is poor; the height of the observed central peak is larger than calculated. This indicates that the transition into the antiferromagnetically ordered state is of first order. In fact, the splitting does not decrease continuously toward $T_{\rm{N}}$ as would be expected for a second-order phase transition.

\begin{figure}
\includegraphics[width=9.5cm,clip]{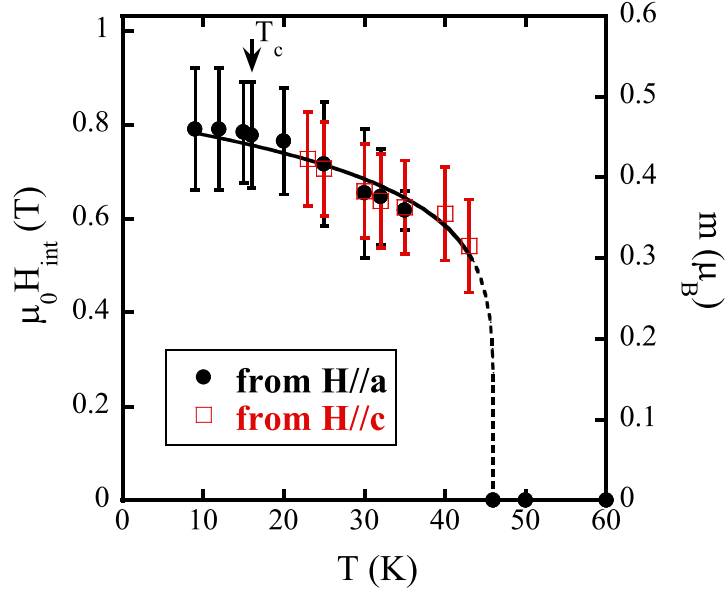}
\caption{\label{fig:Li_Fig3} (Color online) $T$ dependence of the internal magnetic field $H_{\rm{int}}$ (left axis) and the ordered magnetic moment (right axis). The  circles and open squares are deduced from the spectra with $H \parallel a$ and $H \parallel c$, respectively. The  curve is a guide to the eyes.}
\end{figure}

The internal field $H_{\rm{int}}$ can be deduced from the shift of the central peak for $H \parallel a$ and/or the splitting of the  peaks for $H \parallel c$. The temperature dependence of $H_{\rm{int}}$ is shown in Fig. \ref{fig:Li_Fig3} \cite{Supple}. Below $T_{\rm{N}}$ the internal field develops rapidly, reaching to $0.8$ T at $T$=9 K. The saturated internal field is about half that for the undoped parent compound with $T_{\rm{N}} \sim$ 140 K ($H_{\rm int}\sim1.5$ T). For  $H \parallel c$, the signal becomes very weak below $T$=25 K, since the spectrum is spread over a wide frequency range.  For  $H \parallel a$, on the other hand, the uncertainty to calculate $H_{\rm{int}}$ from the peak shift becomes large near $T_{\rm{N}}$.

The right axis of Fig. \ref{fig:Li_Fig3} is the magnitude of the ordered magnetic moment per Fe atom, $m$, which is  deduced from $H_{int}=A_{hf}\cdot m$  by  assuming that the  hyperfine coupling constant $A_{hf}$ is the same as in the undoped compound \cite{KKitagawa}. The estimated moment size at $T$=9 K is about 0.45 $\mu_B$, which is about half that in the undoped compound \cite{WBao}.
At the moment, we cannot rule out the possibility that the hyperfine coupling constant increases upon doping, since the bond length changes upon doping. In that case, the ordered moment can be smaller than displayed in Fig. \ref{fig:Li_Fig3}.

\begin{figure}
\includegraphics[width=10cm,clip]{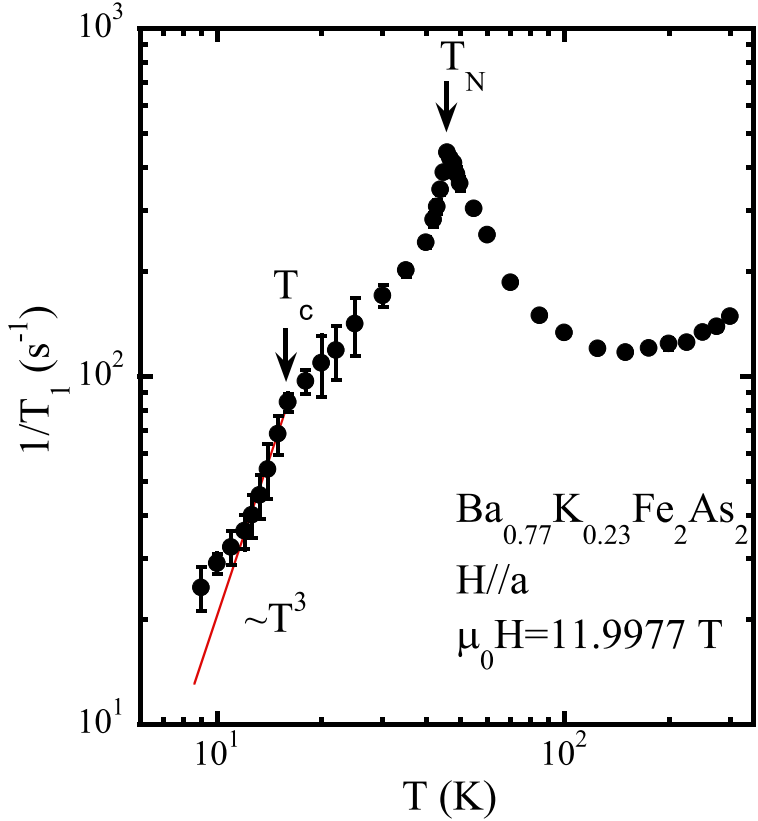}
\caption{\label{fig:Li_Fig4} (Color online) The temperature dependence of the spin-lattice relaxation rate $1/T_{1}$. The straight line indicates the  $1/T_{1} \propto T^3$ relation.}
\end{figure}

Next we discuss the temperature dependence of the  spin lattice relaxation rate $1/T_{1}$ which is measured at the central peak for $H \parallel a$ and plotted in Fig. \ref{fig:Li_Fig4}.   The $1/T_{1}$ shows an upturn with decreasing $T$ and forms a peak  at $T_{\rm{N}}$=46 K, due to a critical slowing down of the magnetic moments.
Below $T_{\rm{N}}$=46 K, $1/T_{1}$ decreases and becomes to be nearly in proportion to $T$ before superconductivity sets in.  Below $T_{\rm{c}}(\mu_{0}H$=12 T)=16 K, $1/T_{1}$ shows another sharp reduction, and follows a  $T^{3}$ relation.
At further low temperatures, the decrease of $1/T_{1}$ becomes gradual.

It should be emphasized that below $T_{\rm{N}}$=46 K, $1/T_{1}$ was measured at the shifted peak that experiences an internal magnetic field. Therefore, the sharp decrease of $1/T_{1}$ below $T_{\rm{c}}$ indicates that the electrons  that are hyperfine coupled to the nuclei produce both the  magnetic order and superconductivity. Our results are clear and direct evidence  that AFM coexist microscopically with superconductivity.
Furthermore, $H_{\rm{int}}$ is not reduced below $T_{\rm{c}}$, as can be seen in Fig. \ref{fig:Li_Fig3}, which suggests that the magnetic order is determined by an energy scale much larger than that of Cooper pairing.

In Ce-based heavy fermion compounds, the same electronic  band derived from Ce-4f$^1$ electrons is responsible for both AFM and SC, so that the ordered magnetic moment is small \cite{Kawasaki,Zheng}. In such a case, AFM and SC may be envisaged   as different sides of a single coin \cite{ZhangSC}.
In the U-based heavy fermion compound UPd$_2$Al$_3$, which is a multi-band system, on the other hand, the situation is more complex. It is believed that
different electron bands  bear respective responsibility for AFM and SC, which allows a large ordered magnetic moment of 0.85 $\mu_B$ to coexist with SC \cite{Sato,Miyake}. The current compound is also a multi-band system, with some orbitals strongly Hund coupled which are more localized and the others more itinerant \cite{Xiang}. It is plausible that  the moderate size of the ordered moment  arises from the more localized $d$ orbitals, so that it can coexist with SC, which is  mainly due to the more itinerant orbitals. Thus, our work demonstrates  the richness of the physics of multiple-orbital electron systems. The microscopic coexistence of AFM and SC in the present system  also suggests that the Fe pnictides can provide another good opportunity to study the  issues such as quantum critical point and associated physics which have been long debated in cuprate high-$T_{\rm{c}}$ superconductors \cite{Sachdev,Broun}.

We note that the property of the superconducting state with the coexisting magnetism is unusual. Namely,
the temperature dependence of $1/T_{1}$ below $T_{\rm{c}}$ is much weaker than in
the optimally doped sample Ba$_{0.68}$K$_{0.32}$Fe$_{2}$As$_{2}$ ($T_{\rm{c}}$=38.5 K), where $1/T_{1}$ follows an exponential decrease down to very low temperatures \cite{Li}. Impurity scattering can hardly explain the difference since both samples have a similar degree of cleanness as evidenced by the similar NMR linewidth ($\sim$83 kHz at $T$=100 K and $\mu_{0}H$=12 T).  We therefore attribute such a weaker $T$ dependence to the coexisting magnetism.
One possibility is that fluctuations of the coexisting magnetic moment contribute significantly to 1/$T_1$ in the superconducting state. This is an un-explored frontier and we hope that our finding will stimulate more theoretical works.
Other possible explanations include two existing theories.
One is the odd-frequency superconducting state proposed for heavy fermions near a quantum critical point which is a gapless state \cite{Fuseya}. The other is a theory proposed for iron pnictides where a nodal superconducting gap is stabilized in the doping region coexisting with magnetic order \cite{Chubukov}.

\begin{figure}
\includegraphics[width=9cm,clip]{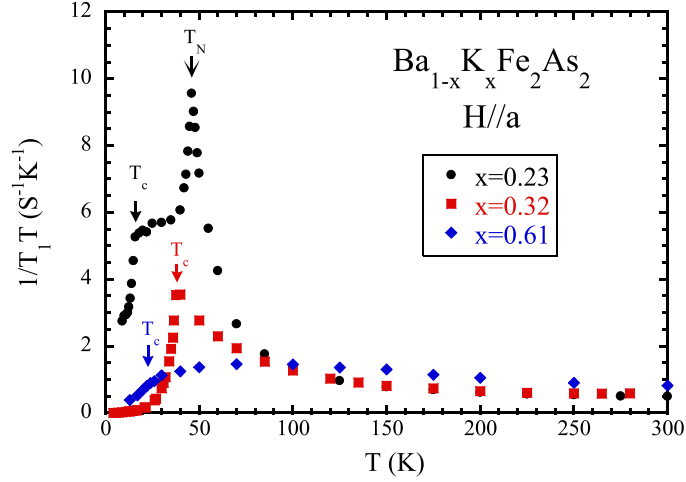}
\caption{\label{fig:Li_Fig5} (Color online) The $T$ dependence of $1/T_{1}T$ of Ba$_{1-x}$K$_{x}$Fe$_{2}$As$_{2}$ with $x$=0.23, 0.32 and 0.61. }
\end{figure}

For completeness,   we show in Fig. \ref{fig:Li_Fig5} the quantity $1/T_{1}T$ as a function of  $T$ for the underdoped ($x$=0.23), optimally doped ($x$=0.32, $T_{\rm{c}}$=38.5 K),\cite{Li} and overdoped ($x$=0.61, $T_{\rm{c}}$=24.5 K) samples.  None of them shows a Korringa relation ($1/T_{1}T=$const.) expected for a conventional metal. The $1/T_{1}T$ increases with decreasing temperature for the underdoped and optimally-doped samples, which is due to the antiferromagnetic spin fluctuations. At high temperatures, the value of $1/T_{1}T$, which is dominated by the density of states (DOS) at the Fermi level,  increases with increasing doping,  which indicates that the DOS increases with increasing doping.

In conclusion, by $^{75}$As NMR measurements on an underdoped single-crystal Ba$_{0.77}$K$_{0.23}$Fe$_{2}$As$_{2}$ with  $T_{\rm{c}}$=16.5 K, we found clear and direct evidence for a microscopic coexistence of antiferromagnetic order and superconductivity. Below  $T_{\rm{N}}$=46 K, the NMR peaks for $H \parallel c$ split and those  for $H \parallel a$ shift to  higher frequencies, which indicates that an internal magnetic field  develops due to the ordered Fe moment lying on the $ab$-plane. The spin lattice relaxation rate $1/T_{1}$ measured at the shifted peak with $H \parallel a$ shows distinct decreases at $T_{\rm{N}}$=46 K and $T_{\rm{c}}(\mu_{0}H$=12 T) = 16 K respectively. Since the nuclei corresponding to the shifted peak experience an internal magnetic field below $T_{\rm{N}}$, our results show unambiguously that the electrons that are hyperfine coupled to the nuclei  produce  both the antiferromagnetic order and form Cooper pairs below $T_{\rm{c}}(\mu_{0}H$=12 T) = 16 K. The superconducting state with the coexisting magnetism is unusual and deserves further studies, in particular theoretically.

We thank K. Miyake, T. Xiang,   R. Fernandes and H. Fukuyama for useful discussions and comments. This work  was supported by National Basic Research Program of China (973 Program) No. 2012CB821402 and No. 2011CBA00109, NSFC Grant No. 11104336, and by CAS. Work at Okayama was supported by  MEXT Grant No.22103004.



\end{document}